\documentclass[%
twocolumn,
superscriptaddress,
 amsmath,amssymb,
 aps,
]{revtex4-1}

\usepackage{graphicx}
\usepackage{color}
\usepackage{dcolumn}
\usepackage{bm}

\usepackage{physics}
\usepackage{cancel}

\def\mean#1{\langle#1\rangle}
\newcommand{\lr}[1]{\left\langle #1 \right\rangle}
\newcommand{\lrabs}[1]{\left| #1 \right|}
\newcommand{\epi}{\varepsilon}
\newcommand{\ie}{i.e.\ }
\newcommand{\eg}{e.g.\ }
\newcommand{\Npop}{N_{\text{pop}}}
\newcommand{\bin}{\Delta t_{\text{bin}}}
\newcommand{\e}[1]{Eq.~(\ref{eq:#1})}
\newcommand{\bi}[1]{Fig.~\ref{fig:#1}}
\newcommand{\D}{\mathcal{D}}

\begin{document}

\title{
Detecting a periodic signal by a population of spiking neurons in the weakly nonlinear response regime}

\author{Maria Schlungbaum}
\affiliation{Physics Department of Humboldt University Berlin}
\affiliation{Bernstein Center for Computational Neuroscience Berlin}

\author{Benjamin Lindner}
\affiliation{Physics Department of Humboldt University Berlin}
\affiliation{Bernstein Center for Computational Neuroscience Berlin}

\date{\today}


\begin{abstract}
Motivated by experimental observations, we investigate a variant of the cocktail party problem: the detection of a weak periodic stimulus in the presence of fluctuations and another periodic stimulus which is stronger than the periodic signal to be detected. Specifically, we study the response of a population of stochastic leaky integrate-and-fire (LIF) neurons to two periodic signals and focus in particular on the question, whether the presence of one of the stimuli can be detected from the population activity. As a detection criterion, we use a simple threshold-crossing of the population activity over a certain time window. We show by means of the receiver operating characteristics (ROC) that the detectability depends only weakly on the time window of observation but rather strongly on the stimulus amplitude. Counterintuitively, the detection of the weak periodic signal can be facilitated by the presence of a strong periodic input current depending on the frequencies of the two signals and on the dynamical regime in which the neurons operate. Beside numerical simulations of the model we present an analytical approximation for the ROC curve that is based on the weakly nonlinear-response theory for a stochastic LIF neuron. We discuss the validity of this approximation as well as the relevance of our results for a detection problem in weakly electric fish.
    
\end{abstract}

\maketitle


\section{Introduction}
The detection of a weak signal in presence of a much stronger signal is an interesting problem that arises in several natural situations for living organisms, most prominently in auditory perception where it is known as the cocktail party problem \cite{McD09}. Detection is complicated by nonlinearity in the sensory apparatus (see \eg \cite{MarHud01b}) and by different noise sources \cite{FaiSel08}, so studying very simple models can help us to better understand this problem. Generally, how neural populations of spiking neurons respond to time-dependent stimuli has been addressed with different theoretical approaches \cite{Kni72a,Ger00,FouHan03,MazPan08,GerKis14,MusGer19,KnoLin22}.

An intriguing example of a signal detection task can be found in the courtship behavior of weakly electric fish. It has been observed that a resident male is able to detect a distant male intruder while courting a female \cite{HenKra18}. This represents an instance of the aforementioned cocktail party problem: a comparatively weak low-frequency signal (the distant intruder) has to be detected in the presence of another time-dependent input, coming from the nearby female, a strong high-frequency stimulus. The detection of this faint signal is a fascinating problem that involves many levels of neural processing and is additionally complicated  by  the movements of the participating fish \cite{HenKra18}, by the change of their frequencies  (known as jamming avoidance response) \cite{EngZup01}, and by different adaptation mechanisms (starting with a pronounced spike-frequency adaptation in the sensory receptor cells, the P-units) \cite{NelXu97}.

Here we take the specific experimental observation of an intruder detection as an inspiration to study a generic detection problem of how one periodic signal can be detected in the presence of another one by a population of stochastically spiking neurons. The generic scheme may also be for other sensory modalities in which concurring periodic signals are present (the sense of hearing would be an obvious case). We would like to stress  that our simple detection scheme is not supposed to account for the observed detection performance (none of the above-mentioned complications is taken into account) but, to the best of our knowledge,  even this simple model has not been properly studied yet. The key questions that we will be interested in here are: Under which conditions is the periodic signal easier to detect - in the presence or the absence of the second (strong periodic) background stimulus? Which role is played by the stochastic firing regime of the neurons in the population (mean- or fluctuation-driven regime) and by the response regime (linear or weakly nonlinear)? Are there specific frequency combinations of the two stimuli that make the weaker signal better detectable?

Assuming a simple detection scheme based on spike counts, we want to explore the roles of linear and nonlinear responses in the stochastic dynamics of the single cell. We hypothesize that the weakly nonlinear response may play a beneficial role in the detection task and that also the presence of the strong periodic background stimulus (i.e. the female courtship input in the above example of weakly electric fish) does not have to be necessarily detrimental for the detection but, on the contrary, may facilitate the detection.

Our paper is organized as follows: first we introduce the model and sketch how we perform the measurement process and how the detector functions. We then present our analytical approximation of the receiver operating characteristic (ROC) which quantifies the detection performance. We investigate the influence of variations of the simulation and detection parameters in the detectability of the periodic signal and focus in particular on the detection time window, the strength of the signal amplitude, and the frequency combinations. We consider two operating regimes of the neuron model, the mean-driven and the excitable regimes and ask in all situations whether the presence of a strong periodic background stimulus can be beneficial for the detection task.


\section{Model and Methods}


\subsection{Population model and single-neuron model}
Inspired by the signal detection problem in the weakly electric fish mentioned in the introduction, we consider a population of spiking neurons that are not connected but driven by a common periodic signal and individual noise. In the example of weakly electric fish, the neurons would correspond to the P-units in the electric fish and the periodic input signal would contain components stemming from the nearby female fish and from the intruder to be detected. The scenario of a population of uncoupled noisy neurons that transmit information of periodic input stimuli is, however, more general and for instance also encountered in the auditory periphery. 

The dynamics of the $i$-th spiking neuron is given by a leaky integrate-and-fire (LIF) model driven by an external signal $s(t)$:
\begin{align}
    \dot{v}_{i}(t) &= -v_{i}(t) + \mu + \epi s(t) + \sqrt{2 D} \xi_{i}(t) \ , \ i = 1, \ldots, \Npop \ .
    \label{eq:lif}
\end{align}
Here $v_{i}(t)$ is the membrane voltage, $\mu$ is the mean input current, $\xi_{i}(t)$ is white Gaussian noise with zero mean $\lr{\xi_{i}(t)} = 0$ and correlation function $\lr{\xi_{i}(t)\xi_{i}(t')} = \delta(t-t')$ and $D$ is the noise intensity. Whenever $v_{i}(t)$ hits the threshold $v_{T}$, a spike is registered for that time and $v_{i}(t)$ is reset to $v_{R}$. In our non-dimensional model, time is measured in units of the membrane time constant, the voltage in multiples of threshold-reset difference, and we set $v_{R} = 0$ and $v_{T} = 1$ (cf. \cite{VilLin09}). In our setup, the sensory stimulus, common to all $\Npop$ units in the population, reads
\begin{align}
    s(t) = a_{s} \cos(\omega_{s} t + \varphi_{s}) + a_{b} \cos(\omega_{b} t + \varphi_{b}) \ .
    \label{eq:two_cosines}
\end{align}
It is given by the sum of two cosine functions with different frequencies {$\omega_{s,b} = 2\pi f_{s,b}$ (we will use both the regular frequencies $f_{s,b}$ and circular frequencies $\omega_{s,b}$), relative amplitudes $a_{s,b}$ and phase offsets $\varphi_{s,b}$; the total signal $s(t)$ enters the dynamics scaled by a global amplitude $\epi$. The two terms represent the total stimulus, consisting of a strong background stimulus  $a_{b} \cos(\omega_{b} t + \varphi_{b})$ and the weak signal $a_{s} \cos(\omega_{s} t + \varphi_{s})$ the presence of which has to be detected from the output of the population. In the example of the weakly electric fish, these two effective signals emerge from a beating pattern of the considered fish's own electric organ discharge and those of the intruder fish (leading to $a_{s} \cos(\omega_{s} t + \varphi_{s})$) and of the female fish  (leading to $a_{b} \cos(\omega_{b} t + \varphi_{b})$) \cite{BenLon05}. We will consider situations in which the weak (intruder) signal is absent ($a_{s} = 0$) or present ($a_{s} > 0$) and will ask how the presence of this signal  can be detected. We will also inspect how the detectability of the signal  depends on the presence ($a_{b} > 0$) or absence ($a_{b} = 0$) of the background periodic stimulus.

For the numerical simulations of \e{lif} we use the Euler-Maruyama method, operating in discrete time steps $t = t_{0} + \ell \Delta t$
\begin{align}
    v_{i}\bigl(t_{0} + (\ell+1) \Delta t \bigr) =& v_{i}(t_{0} + \ell \Delta t) \pqty{1 - \Delta t} + \mu \Delta t \nonumber \\
    & + \epi s(t_{0} + \ell \Delta t) \Delta t + \sqrt{2D \Delta t} \vartheta_{\ell} \ .
    \label{eq:lif_numeric}
\end{align}
Here $\vartheta_{\ell}$ are independent Gaussian numbers drawn from a normal distribution with zero mean $\mean{\vartheta_{\ell}} = 0$ and unit variance $\lr{\vartheta_{\ell} \vartheta_{k}} = \delta_{\ell k}$. For all numerical examples in this paper, we use an integration time step of $\Delta t = 10^{-3}$.

We consider a neuron population of $\Npop = 10^{3}$ in all simulations. We analyze the time-dependent count statistics of the population for the different signal combinations in order to test a specific idea how the presence of the weak periodic signal may be detected (for details see below). To determine the time-dependent spike count $N(t)$, we discretize the time axis into bins $\bin = 0.05$ (significantly larger than our integration time step) and count all the spikes fired by all neurons within this bin:
\begin{align}
    N(t) = \sum_{i=1}^{\Npop} N_{i}[t, t + \bin] \ .
\end{align}
Dividing this by the size of $\bin$
and an additional trial average yields an approximation of the instantaneous firing rate
\begin{align}
    r(t) \approx \sum_{i=1}^{\Npop} \frac{\lr{N_{i}[t, t + \bin]}_{\xi_{i}}}{\bin \Npop} \ ,
    \label{eq:rate_count_relation}
\end{align}
where $\lr{\cdot}_{\xi_{i}}$ indicates an ensemble average over realizations of the intrinsic noise $\xi_{i}(t)$ (the common signal is always the same in all these realizations).


\subsection{Detection Process}
Our approach here is similar in spirit to the recent study by Bernardi and Lindner \cite{BerLin20} of the detection of a static signal embedded in an Ornstein-Uhlenbeck process; however, there are also important differences (see below).

Specifically, we will analyze two time series for different experiments, i.e., spike count modulations $N(t)$ in the presence or absence of the signal. We will carry out this numerical experiment for the two distinct situations when the strong periodic background signal is present ($a_{b} = 1$ in \e{two_cosines}) or not ($a_{b} = 0$ in \e{two_cosines}). We measure the time-dependent counts in a very long time window that is split into $N_{T}$ smaller detection windows $T_{j}$ of length $T=K\bin$ and short pauses of length $\Delta t_{\text{off}}$ cf. \bi{fig1}. The $N_{T}$ time windows serve as trials - this is somewhat different to the procedure in \cite{BerLin20}, where trials result from the repetition of the same experiment. At the same time the averaging over subsequent time windows implies an automatic averaging over the initial phases of the periodic signals (as long as the time window is not a multiple of one the driving signal's period, a non-generic case that we exclude in the following).

\begin{figure}[h]
    \centering
    \includegraphics[width=0.49\textwidth]{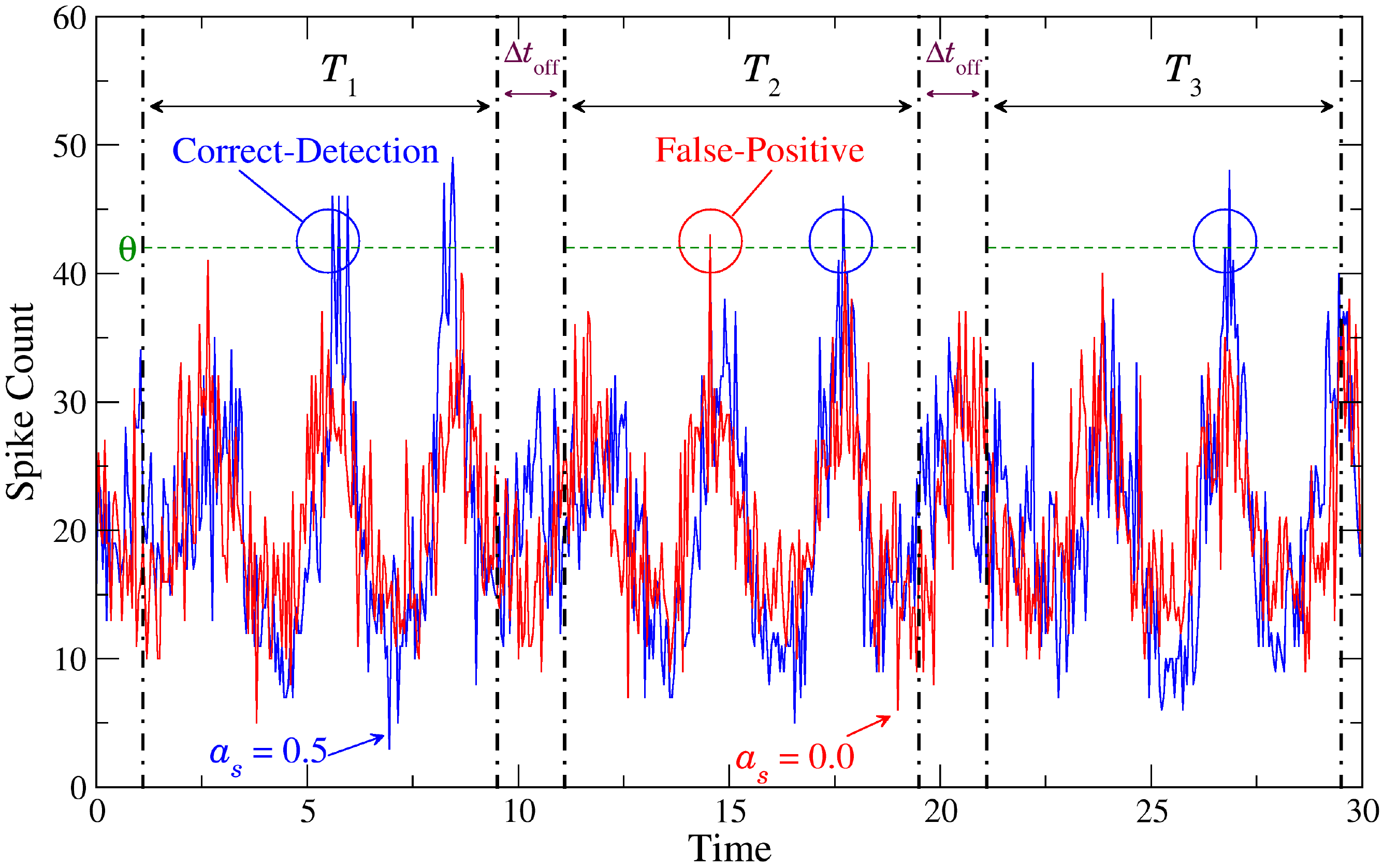}
    \caption{Illustration of the measurement process. One spike count modulation $N(t)$ in presence of the signal (blue) and one in absence of the signal (red) are shown as well as the first three detection windows. In all visible trials, the chosen threshold $\theta$ is exceeded by the blue trajectory implying a registration of a correct-detection event for these trials. The red trajectory reaches $\theta$ only in $T_{2}$, \ie we record a false-positive event for $T_{2}$ but not for $T_{1}$ or $T_{3}$. Remaining parameters: $a_{b} = 1.0$, $f_{s} = 0.1$, $f_{b} = 0.33$, $\epi = 0.05$, $\bin = 0.05$
    }
    \label{fig:fig1}
\end{figure}

We assume that the detection of an event takes place whenever the spike count of the population crosses a threshold $\theta$ (green dashed horizontal line in \bi{fig1}). In two distinct numerical simulations of our population model the count in the presence (blue) and in the absence of the signal (red) is measured, respectively. When the blue time series crosses $\theta$ at least once within the corresponding time window, a correct-detection is registered for that trial. In analogy to this, a false-positive event is recorded when the red time series exceeds the threshold at least once. The correct-detection (CD) and false-positive (FP) rates are then obtained by averaging over all $N_{T}$ trials. Varying the threshold yields the two rates as functions of $\theta$.

In \bi{fig2} a few examples for the FP and CD rates vs threshold $\theta$ for different values of the signal amplitude $a_{s}$ are shown in (a) together with the corresponding ROC curves in (b). The latter are obtained by plotting the CD rate as a function of the FP rate. A low $\theta$ corresponds to a very high detector sensitivity and is indicated by the upper right corner in \bi{fig2}(b), a high $\theta$ is represented by the lower left range. The example curves are taken for $a_{b} = 1$, which means in presence of the strong periodic background stimulus (i.e. presence of the female fish in the courtship example).

\begin{figure}[h]
    \centering
    \includegraphics[width=0.49\textwidth]{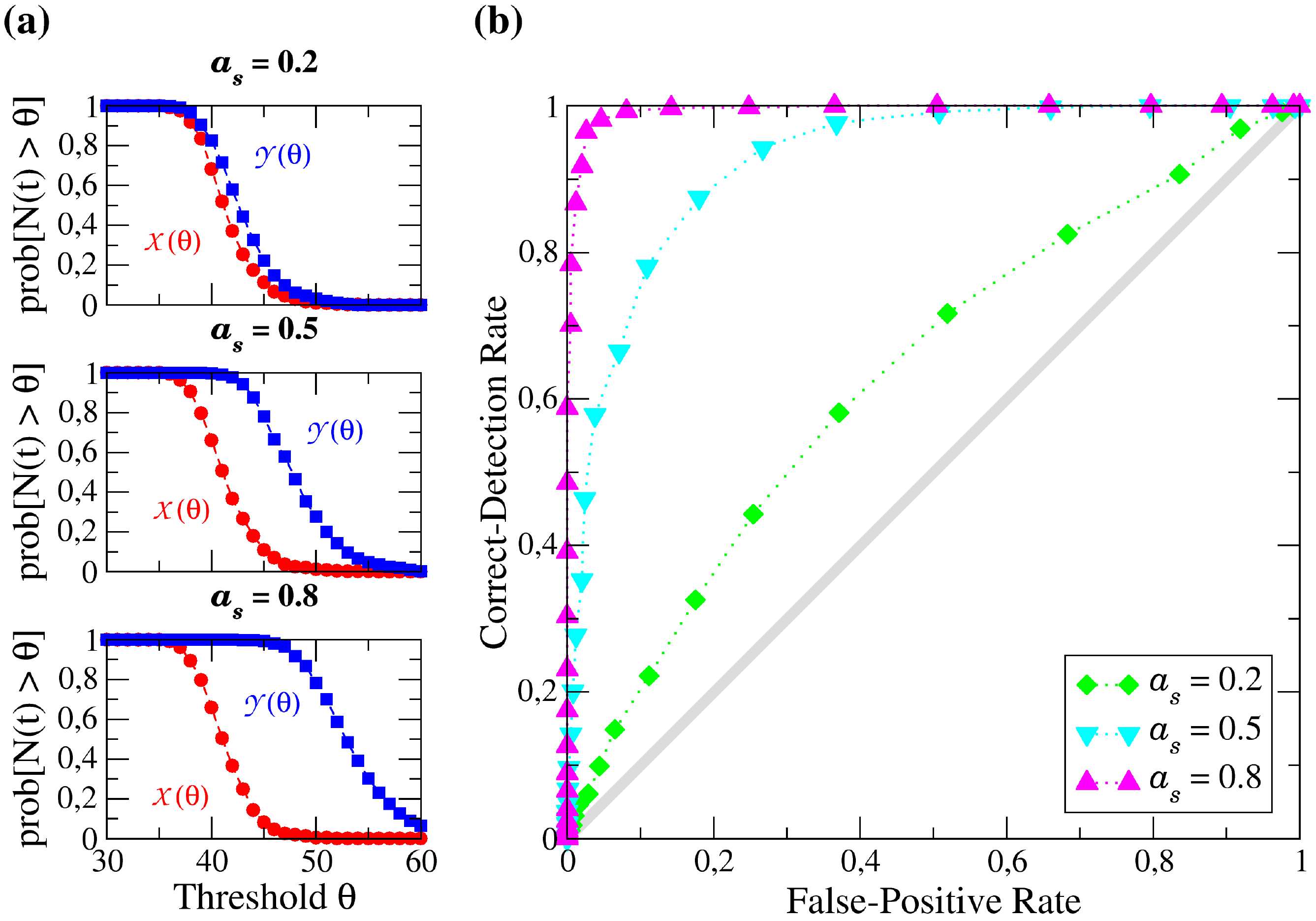}
    \caption{(a) FP rates $\mathcal{X}(\theta)$ and CD rates $\mathcal{Y}(\theta)$ obtained from simulations for different values of the signal amplitudes as indicated in each subplot title, (b) corresponding ROC curves. Remaining parameters: $a_{b} = 1$, same as in \bi{fig1}
    }
    \label{fig:fig2}
\end{figure}

We first focus on the CD (blue) and FP (red) rates as functions of the threshold $\theta$ in \bi{fig2}(a). At small $\theta$ the probability of the measured spike count $N(t)$ to be above the threshold at least once within the time window is close to one and decreases with increasing $\theta$ towards zero. For a weak signal, $a_{s} = 0.20$, the CD and FP rates are very close to each other implying that the detection of the signal is difficult. Boosting the signal amplitude leads to a higher spike count, thus the CD rate starts to decrease for higher threshold values while the FP rate remains the same. As a consequence, the horizontal distance between the CD and FP rates is increasing for higher signal amplitudes and detection becomes a simpler task as can be expected. The improved detectability for larger $a_{s}$ is also apparent in the ROC curves in \bi{fig2}(b). For a weak signal amplitude ($a_{s} = 0.2$, green), the distance to the diagonal (grey solid line, representing chance level) is small. Enlarging the signal amplitude yields a growing distance from the diagonal. In the present examples the ROC curve of the signal amplitude $a_{s} = 0.8$ (pink) is already close to that of an ideal detector, \ie a detector with a $100\%$ CD rate at all FP rates.

\subsection{Analytical approximation of the ROC}
To derive an analytical description of the ROC curve, we make two assumptions for the count in a short time bin ($\bin \ll \mean{I}$, where $I$ denotes the interspike interval of a single stochastic LIF neuron):
\begin{enumerate}
    \item The spike count distribution follows a Poisson distribution $P_{\mean{N_{k}}} (N)$;
    \item spike counts in each time bin are independent.
\end{enumerate}
The first assumption is well justified for a sufficiently small time bin (see \eg \cite{MidCha03}). The second assumption is certainly an approximation, as the sum of independent non-Poissionian spike trains do not converge to a Poisson process \cite{Lin06}. Assuming a Poisson count statistics has the great advantage that we just need to know the mean spike count $\mean{N_{k}}$ to determine the distribution completely. The basic idea is identical to the procedure in the numerical case: we want to estimate the probability to be at least once in the detection window above a certain threshold $\theta$.

We first determine the probability for the spike count $N$ in a time bin $\bin$ to be below or at most at the threshold $\theta$. This is given by the sum over all possible values of $N \in \mathbb{N}$ up to $\lfloor\theta\rfloor$, the largest integer smaller than or equal to $\theta$: 
\begin{align}
    \text{prob} \pqty{N \leq \theta} &= \sum_{N = 0}^{\lfloor\theta\rfloor} P_{\mean{N_{k}}} (N) = e^{-\mean{N_{k}}} \sum_{N = 0}^{\lfloor\theta\rfloor} \frac{\mean{N_{k}}^{N}}{N!} \nonumber \\
    &= \frac{\Gamma \pqty{1 + \lfloor\theta\rfloor, \mean{N_{k}}}}{\lfloor\theta\rfloor !} \approx \frac{\Gamma \pqty{1 + \theta, \mean{N_{k}}}}{\Gamma(1 + \theta)} \ .
    \label{eq:problowtheta}
\end{align}
In the last line we used the incomplete Gamma function $\Gamma(a,x) = \int_{x}^{\infty} \text{d} t \ t^{a-1} e^{-t}$ \cite{AbrSte70} and furthermore provided in the last step an approximate expression that interpolates between integer values of $\theta$ (at the latter, the two expressions coincide). In the following we will use for simplicity only the latter expression.

To estimate the probability $p(\theta, T)$ to be not even once above $\theta$ in the $j$-th detection window $T_{j}$, we multiply the probabilities \e{problowtheta} from all bins, exploiting the assumption of statistical independence that holds true for a Poisson process:
\begin{align}
    p(\theta, T) &= \prod_{k = 0}^{K-1} \frac{\Gamma \pqty{1 + \theta, \mean{N(t_{j;k})}}}{\Gamma \pqty{1 + \theta}} \ .
    \label{eq:problowthetafull}
\end{align}
Here we have used the population spike counts at times $t_{j,k} = j(T + \Delta t_{\text{off}})+k\bin$.

The FP $\mathcal{X}(\theta, 0, T)$ and CD $\mathcal{Y}(\theta, a_{s}, T)$ rates are then given by
\begin{align}
    \mathcal{X}(\theta, 0, T) &= \frac{1}{N_{T}} \sum_{j=0}^{N_{T}-1} \left(1 - \Biggl. \right. \nonumber \\
    &\qquad \left. \prod_{k=0}^{K-1} \frac{\Gamma \pqty{1 + \theta, \lr{N(t_{j,k}; a_{s} = 0)}}}{\Gamma \pqty{1 + \theta}}\right)\ ,
    \label{eq:fpr}
\end{align}
\begin{align}
    \mathcal{Y}(\theta, a_{s}, T) &= \frac{1}{N_{T}} \sum_{j=0}^{N_{T}-1} \left(1 - \Biggr. \right. \nonumber \\
    &\qquad \left. \prod_{k=0}^{K-1} \frac{\Gamma \pqty{1 + \theta, \lr{N(t_{j,k}; a_{s} > 0)}}}{\Gamma \pqty{1 + \theta}}\right)\ ,
    \label{eq:cdr}
\end{align}
where we indicated the explicit dependence of the spike count on the signal amplitude by a parametric argument. Furthermore, the above formulas also include the trial average over the detection windows $T_{j}$ (sum over $j$).

With the obtained expression we can proceed in two different ways. Firstly, we can measure the time-dependent mean spike count in simulations and use these data in \e{fpr} and \e{cdr}. This will be referred to as a semi-analytical theory in the following as it still requires some numerical simulations. Secondly, we can approximate the mean spike count using linear and nonlinear response theory for stochastic integrate-and-fire neurons that are driven by periodic signals \cite{FouBru02,VorLin17}. 
In the latter case, we will use \e{rate_count_relation} to estimate the mean spike count via
\begin{align}
	\mean{N(t_{j;k})} \approx r(t_{j;k}) \bin \Npop\ .
\end{align}

The instantaneous firing rate in the weakly nonlinear regime (neglecting higher than second-order terms in $\epi$) can be approximated by \cite{VorLin17}
\begin{widetext}
\begin{align}
    r(t) &\approx r_{0} + \frac{\epi^{2} a_{s}^{2}}{2} \chi_{2} \pqty{\omega_{s}, -\omega_{s}} + \frac{\epi^{2} a_{b}^{2}}{2} \chi_{2} \pqty{\omega_{b}, -\omega_{b}} \nonumber \\
    &\quad + \epi \Bigl[ a_{s} \lrabs{\chi_{1}(\omega_{s})} \cos \bigl( \omega_{s} t + \varphi_{s} - \phi_{1}(\omega_{s}) \bigr) + a_{b} \lrabs{\chi_{1}(\omega_{b})} \cos \bigl( \omega_{b} t + \varphi_{b} - \phi_{1}(\omega_{b}) \bigr) \Bigr]_{\text{LR}} \nonumber \\
    &\quad + \frac{\epi^{2}}{2} \Bigl[ a_{s}^{2} \lrabs{\chi_{2} \pqty{\omega_{s}, \omega_{s}}} \cos \bigl( 2 \omega_{s} t + 2 \varphi_{s} - \phi_{2} \pqty{\omega_{s}, \omega_{s}} \bigr) + a_{b}^{2} \lrabs{\chi_{2} \pqty{ \omega_{b}, \omega_{b} }} \cos \bigl( 2\omega_{b} t + 2\varphi_{b} - \phi_{2} \pqty{\omega_{b}, \omega_{b}} \bigr) \Bigr]_{\text{HH}} \nonumber \\
    &\quad + \epi^{2} a_{s} a_{b} \Bigl[ \lrabs{\chi_{2} \pqty{\omega_{s}, \omega_{b}}} \cos \bigl( \pqty{\omega_{s} + \omega_{b}}t + \varphi_{s} + \varphi_{b} - \phi_{2} \pqty{\omega_{s}, \omega_{b}} \bigr) \Bigl. \nonumber \\
    &\qquad \qquad \qquad \Bigr. + \lrabs{\chi_{2} \pqty{\omega_{s}, -\omega_{b}}} \cos \bigl( \pqty{\omega_{s} - \omega_{b}}t + \varphi_{s} - \varphi_{b} - \phi_{2} \pqty{\omega_{s}, -\omega_{b}} \bigr) \Bigr]_{\text{MR}} \ .
    \label{eq:rate_ana}
\end{align}
\end{widetext}
Here we have included initial phases $\varphi_{s,b}$ for both periodic signals. The indices in the above expression indicate distinct contributions to the response: the steady state with an index 0, the linear response (LR), the higher harmonics of the periodic driving (HH) and the mixed response (MR) that emerges because of the simultaneous presence of two signals (see \cite{VorLin17} for further discussion). All the firing and response characteristics for the white-noise driven LIF model, $r_{0}, \chi_{1}(\omega), \chi_{2}(\omega_{1},\omega_{2})$, are given in the appendix \ref{sec:app}.

The FP $\mathcal{X}_{\text{ana}}(\theta,0,T)$ and CD $\mathcal{Y}_{\text{ana}}(\theta,a_{s},T)$ rates for the analytical theory can then be expressed by
\begin{align}
    \mathcal{X}_{\text{ana}}&(\theta, 0, T) = \frac{1}{N_{T}} \sum_{j=0}^{N_{T}-1} \left(1 - \Biggl. \right. \nonumber \\
    &\quad \left. \prod_{k=0}^{K-1} \frac{\Gamma \pqty{1 + \theta, r(t_{j,k}; a_{s} = 0) \bin \Npop}}{\Gamma \pqty{1 + \theta}}\right)\ ,
    \label{eq:fpr_ana} \\
    \mathcal{Y}_{\text{ana}}&(\theta, a_{s}, T) = \frac{1}{N_{T}} \sum_{j=0}^{N_{T}-1} \left(1 - \Biggr. \right. \nonumber \\
    &\quad \left. \prod_{k=0}^{K-1} \frac{\Gamma \pqty{1 + \theta, r(t_{j,k}; a_{s} > 0) \bin \Npop}}{\Gamma \pqty{1 + \theta}}\right)\ .
    \label{eq:cdr_ana}
\end{align}

The FP rate in the absence of the strong periodic background stimulus ($a_{b} = 0$) can be simplified: in this case we also have for the signal (intruder) amplitude $a_{s} = 0$, so the instantaneous firing rate in \e{rate_ana} reduces to $r(t)\approx r_{0}$ and we obtain
\begin{align}
	\mathcal{X}_{\text{ana}}&(\theta, T) = 1 - \left(\frac{\Gamma \pqty{1 + \theta, r_{0} \bin \Npop}}{\Gamma \pqty{1 + \theta}}\right)^{K}\ .
    \label{eq:fpr_without_female}
\end{align}


\section{Results}
In the following we investigate the signal detection task in two very different parameter regimes of the LIF neurons: neurons of the population are either mean-driven ($\mu = 1.1 > v_{T}$, $D=0.001$) or in an excitable regime ($\mu = 0.9 < v_{T}$, $D = 0.005$). We will see that the weakly nonlinear response will have a very different impact in the two regimes. We consider variations of the detection and signal parameters for both regimes. We are particularly interested in how the presence of the strong background stimulus  affects the detectability of the intruder. Simulations were performed for a total number of $\Npop = 10^{3}$ LIF neurons and to create the ROC curves $N_{T} = 10^{3}$ detection windows have been used.


\subsection{Change of the detection time window}
Firstly, we would like to study the impact of the detection time window $T$ in the excitable regime. \bi{fig3} shows the ROC curves for a relatively weak signal ($a_{s} = 0.2$), in presence (\bi{fig3}(a), $a_{b} = 1$) and absence (\bi{fig3}(b), $a_{b} = 0$) of the strong periodic background stimulus. \bi{fig3}(c) and (d) show the differences of the CD and FP rates as a function of the FP rate, which is referred to as the \emph{effect size} \cite{HouBre08,BerDor21}. The different values of $T$ are given in multiples of the period of the  signal $T_{s}$ and are represented by different symbols and colors as indicated in the legend. The solid lines represent the analytical theory for the ROC curves (\e{cdr_ana} plotted vs \e{fpr_ana}). Three observations can be made. First of all, for the parameters chosen, the theory is in good agreement with the simulation results. Secondly, the effect of increasing the time window is very weak: a tenfold increase in the detection time window does not even lead to a doubling of the effect size, \ie in the detectability of the intruder. Thirdly, there is not much of a difference in the detectability introduced by the presence of the background stimulus.

\begin{figure}[h]
	\centering
	\includegraphics[width=0.48\textwidth]{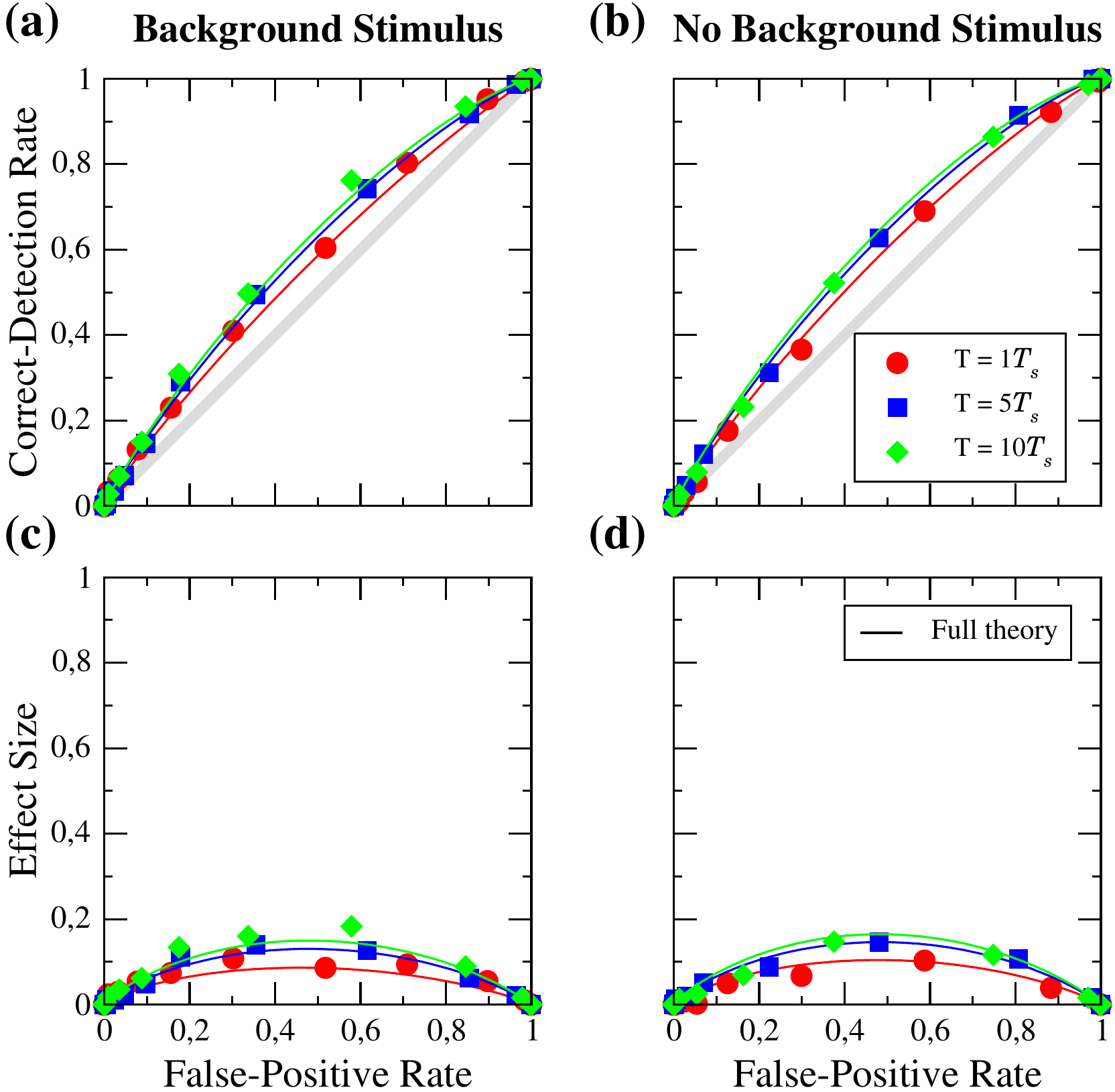}
	\caption{Excitable regime: ROC curves for different lengths of the detection window $T$ in presence (a) and absence (b) of the background stimulus. The length is given in multiples of the period of the  signal $T_{s}$ as indicated in the legend. Symbols are numerical simulations, solid lines represent the analytical theory of the ROC curves. (c) and (d) show the effect size as a function of the FP rate. Parameters: $f_{s} = 0.1$, $f_{b} = 0.33$, $a_{s} = 0.2$}
	\label{fig:fig3}
\end{figure}

Next we look at the same statistics for a larger signal amplitude (cf. \bi{fig4}); the general effect of $a_{s}$ will be inspected in the next subsection. The effect size is generally larger than before, \ie the ROC curves are further away from the diagonal (see panels (a), (b)). The full analytical theory still works and the effect of enlarging the time window is still weak. If we compare the detection in the presence and in the absence of the background stimulus, we find that its presence  can diminish the detection performance slightly. In panel (e) we show the FP and CD rates as functions of the threshold in the absence and the presence of the background stimulus for the detection window of $T=5T_{s}$; the main effect of the background stimulus is to shift the rates to higher thresholds, which has no effect on the ROC curves.

\begin{figure}[h]
	\centering
	\includegraphics[width=0.48\textwidth]{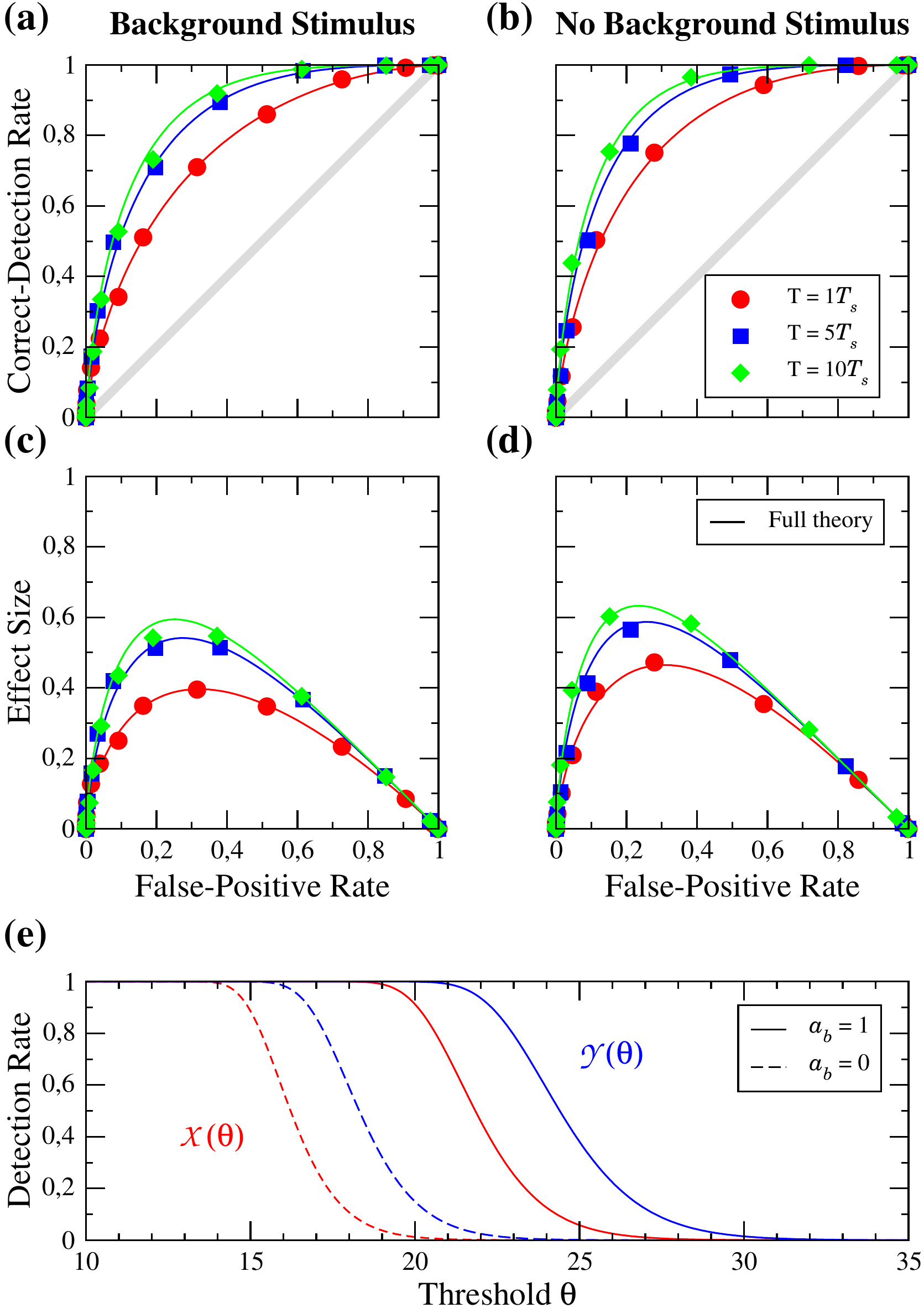}
	\caption{(a)-(d) same as \bi{fig3} but for an signal amplitude of $a_{s} = 0.5$, (e) FP (red) and CD (blue) rates in presence (solid lines) and absence (dashed lines) of the background stimulus.}
	\label{fig:fig4}
\end{figure}

We now turn to the mean-driven regime. The ROC curves and corresponding effect sizes in presence and absence of the strong background stimulus for a signal amplitude of $a_{s} = 0.2$ are shown in \bi{fig5}. In marked contrast to the excitable case, we find that the detector benefits from the presence of the background stimulus; there is a significant boost in the detectability of the weak periodic signal going from panel (d) to panel (c). Indeed, without the background stimulus, the ROC curves are practically on the diagonal and detection is nearly impossible.

\begin{figure}[h]
    \centering
    \includegraphics[width=0.48\textwidth]{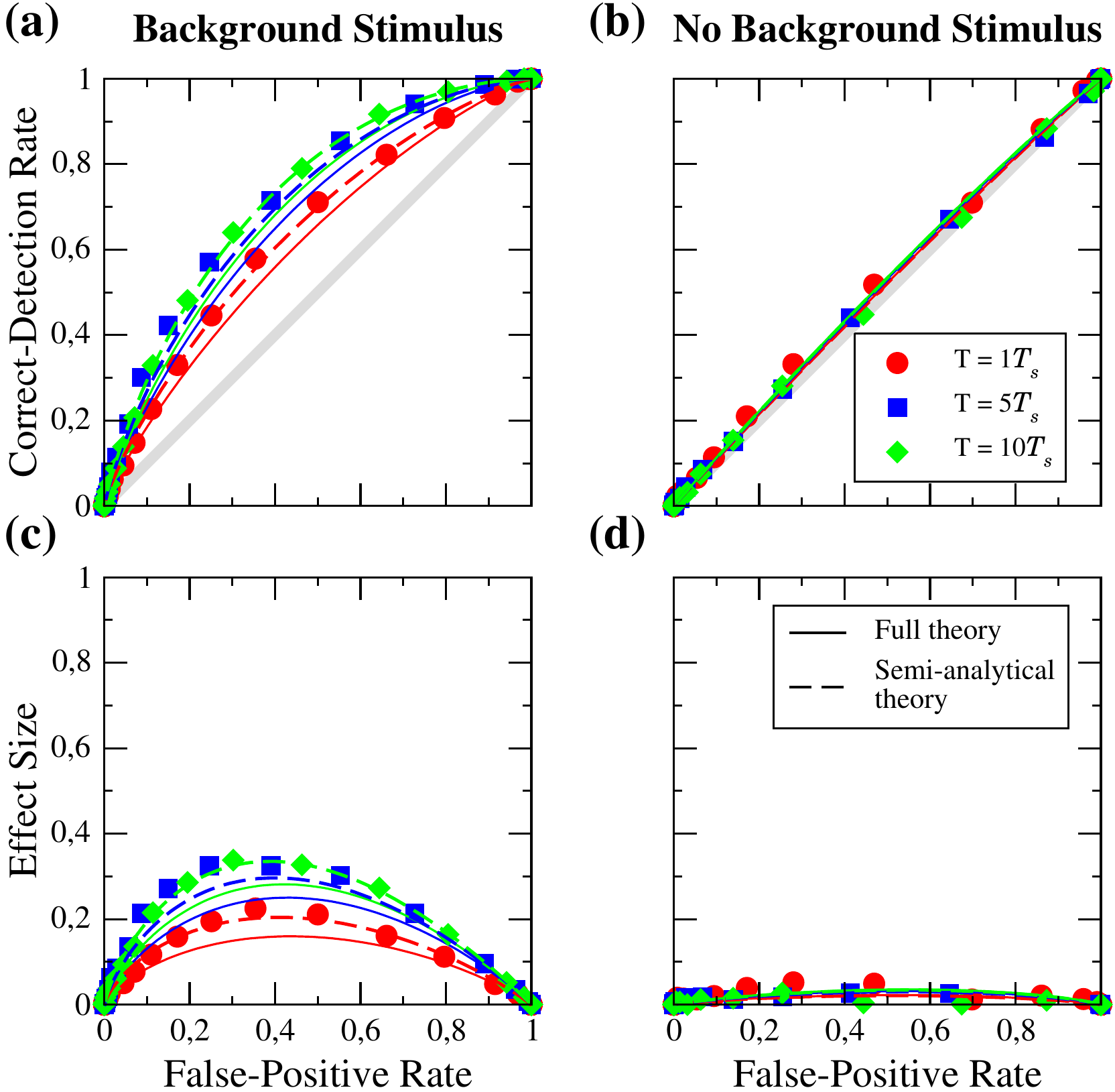}
    \caption{Mean-driven regime: ROC curves for different lengths of the detection window $T$ in presence (a) and absence (b) of the background stimulus. The length is given in multiples of the period of the  signal $T_{s}$ as indicated in the legend. Symbols are numerical simulations, solid lines represent the analytical theory, dashed lines the semi-analytical theory of the ROC curves. (c) and (d) show the effect size as a function of the FP rate. Parameters: $f_{s} = 0.1$, $f_{b} = 0.33$, $a_{s} = 0.2$}
    \label{fig:fig5}
\end{figure}

Like in the excitable regime, the effect of enlarging the detection time window is weak. Furthermore, we notice that the analytical theory (solid lines) differs somewhat from the simulation results. This begs the question which part of our approximate calculation is responsible for this deviation. In order to answer this, we have determined by many simulations of the same periodic stimulus the time-dependent mean spike count that is the crucial input to the semi-analytical theory, \e{fpr}, \e{cdr}. If we use these simulated mean count data (dashed lines in \bi{fig5}) the agreement is again very good. This means that the noticeable deviations of the full theory are due to the limitations of the weakly nonlinear response theory: taking only the terms up to the second-order in $\epi$ does not reproduce the firing rate correctly. To obtain a better approximation of the firing rate and therefore also a good agreement of the analytical theory with the simulation results, one has to take higher order terms into account (see \cite{FraRam22} for such a computation).

\vspace{0.06cm}

Next we look at the effect of using a larger signal amplitude in the mean-driven regime (cf. \bi{fig6}). As in the excitable regime, the effect size is generally larger than before. We again find that the analytical theory differs noticeable from the simulation results but that the use of the numerically determined mean spike counts restores a good agreement. We also note that in the absence of the background stimulus our full analytical theory works well.

\vspace{0.06cm}

\begin{figure}[h]
    \centering
    \includegraphics[width=0.48\textwidth]{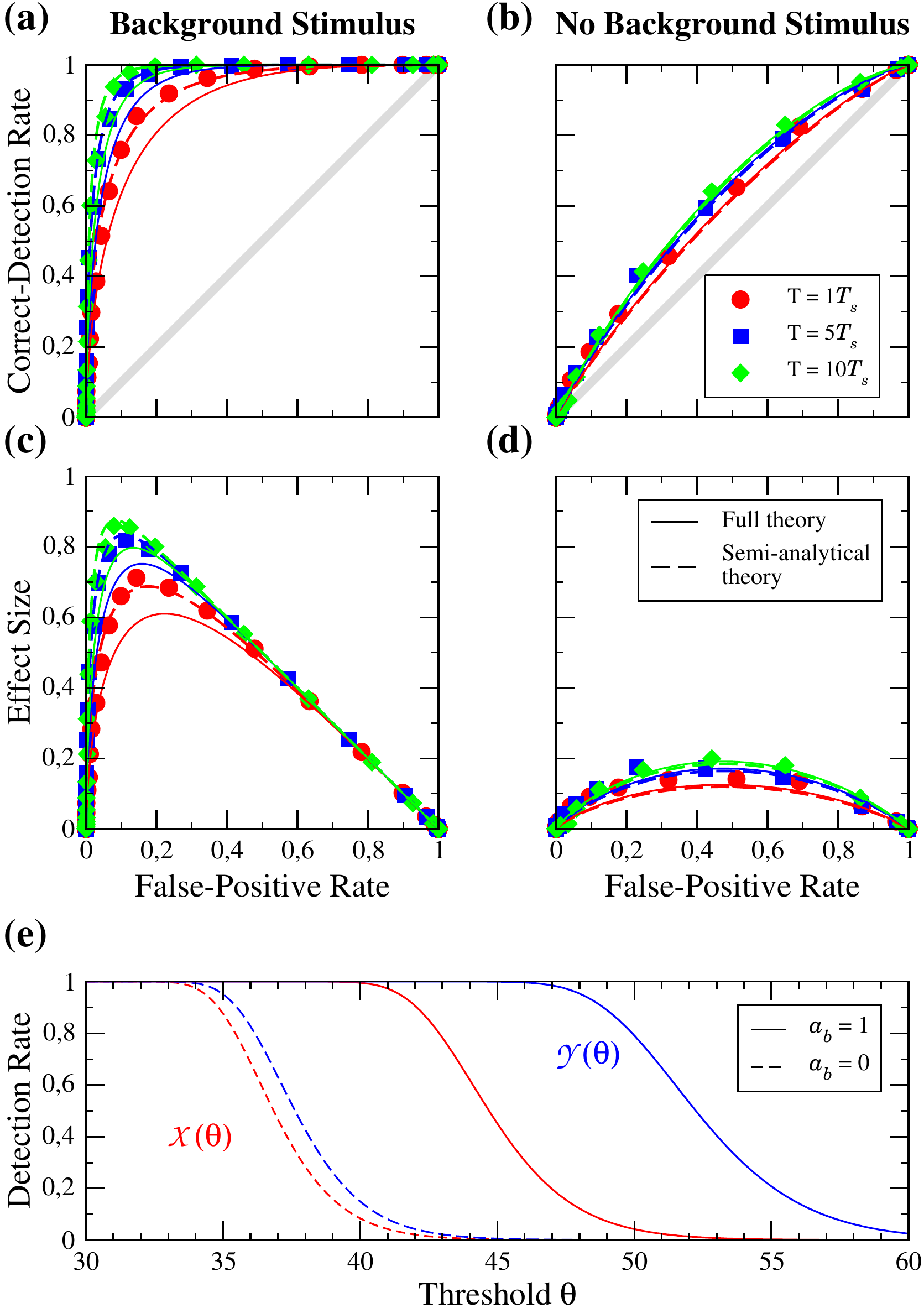}
    \caption{(a)-(d) same as \bi{fig5} but for an signal amplitude of $a_{s} = 0.5$, (e) FP (red) and CD (blue) rates in presence (solid lines) and absence (dashed lines) of the background stimulus.}
    \label{fig:fig6}
\end{figure}

Additionally, we show in \bi{fig6}(e) the FP and CD rates as functions of the threshold. In contrast to the excitable case, in the presence of the background stimulus ($a_{b} > 0$) the two rates are clearly further apart which improves detectability.


\subsection{Dependence on the signal amplitude}
The detection problem is, obviously, particularly interesting for weak to moderate signal amplitudes, which we now explore in more detail. 
We again start with the excitable regime and consider in \bi{fig7} the ROC curves for a range of amplitudes.

\begin{figure}[h]
    \centering
    \includegraphics[width=0.48\textwidth]{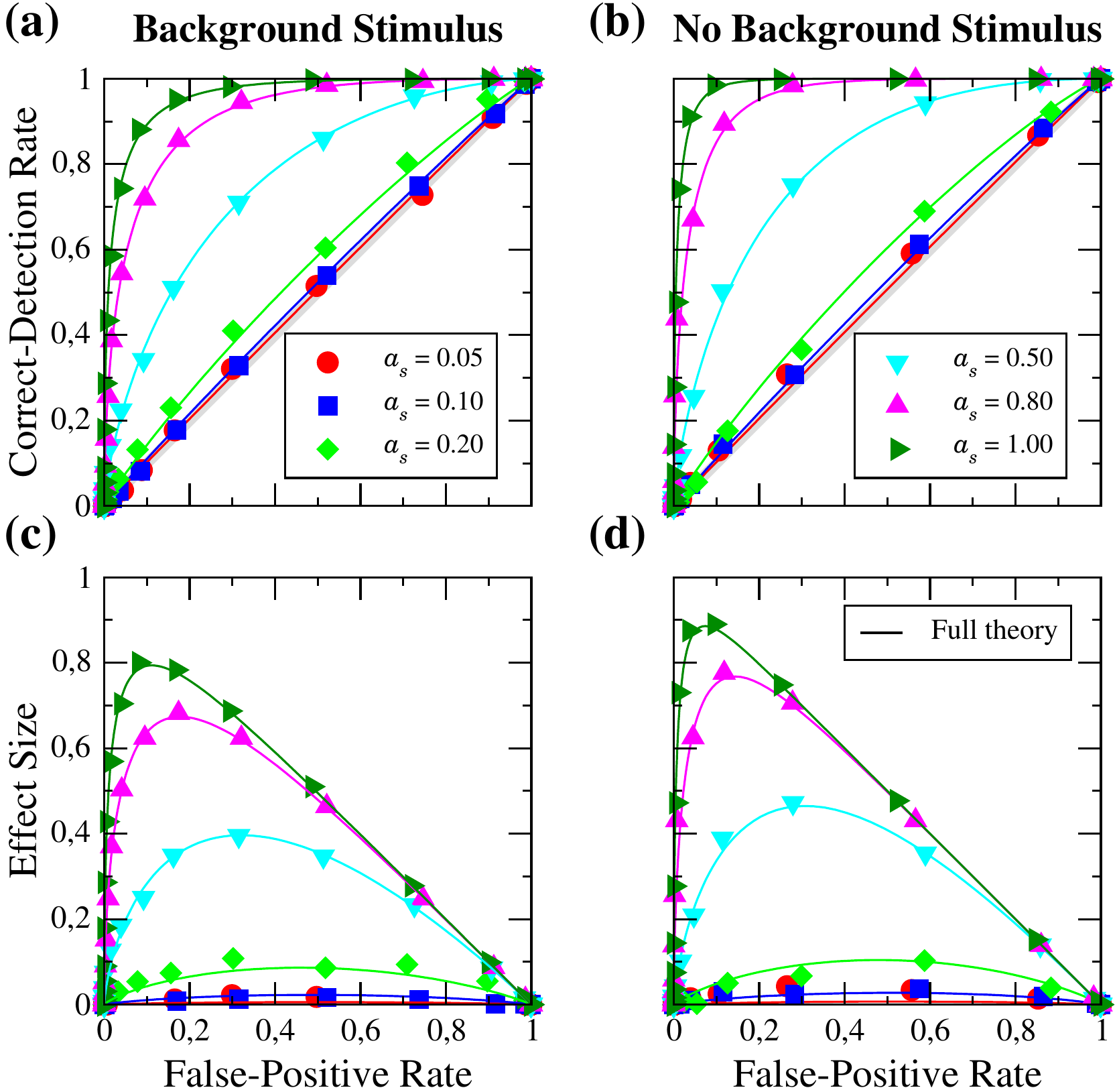}
    \caption{Excitable regime: ROC curves for different strengths of the signal amplitude $a_{s}$ in presence (a) and absence (b) of the background stimulus. (c) and (d) show the effect size as a function of the FP rate. Parameters: $f_{s} = 0.1$, $f_{b} = 0.33$, $T = 10$}
    \label{fig:fig7}
\end{figure}

Referring to our three observations of the first subsection, we firstly notice that the full analytical theory is still in good agreement with our simulation results for all values of $a_{s}$ in presence and absence of the background stimulus . Secondly, increasing the signal  amplitude has a strong effect on the detectability, which is not surprising. But in contrast to enlarging the detection time window, a doubling of the value of $a_{s}$ may also lead to a doubling in the effect size in both cases (see panels (c) and (d)). Thirdly, for weak signal  amplitudes ($a_{s} \le 0.2$) we observe again not much of a difference induced by the presence of the background stimulus . For relatively strong values of $a_{s}$, \ie $a_{s} \ge 0.5$, the detectability of the signal  seems to be better in the absence of the background stimulus , indicated by an up to $10\%$ higher effect size.

In the mean-driven regime, cf. \bi{fig8} we have a clear beneficial role of the background stimulus  in the detection task. The deviations of the analytical theory to our simulation results in presence of the background stimulus  are small and using the numerically determined mean spike count (semi-analytical approach) leads again to an excellent agreement.

\begin{figure}[]
    \centering
    \includegraphics[width=0.48\textwidth]{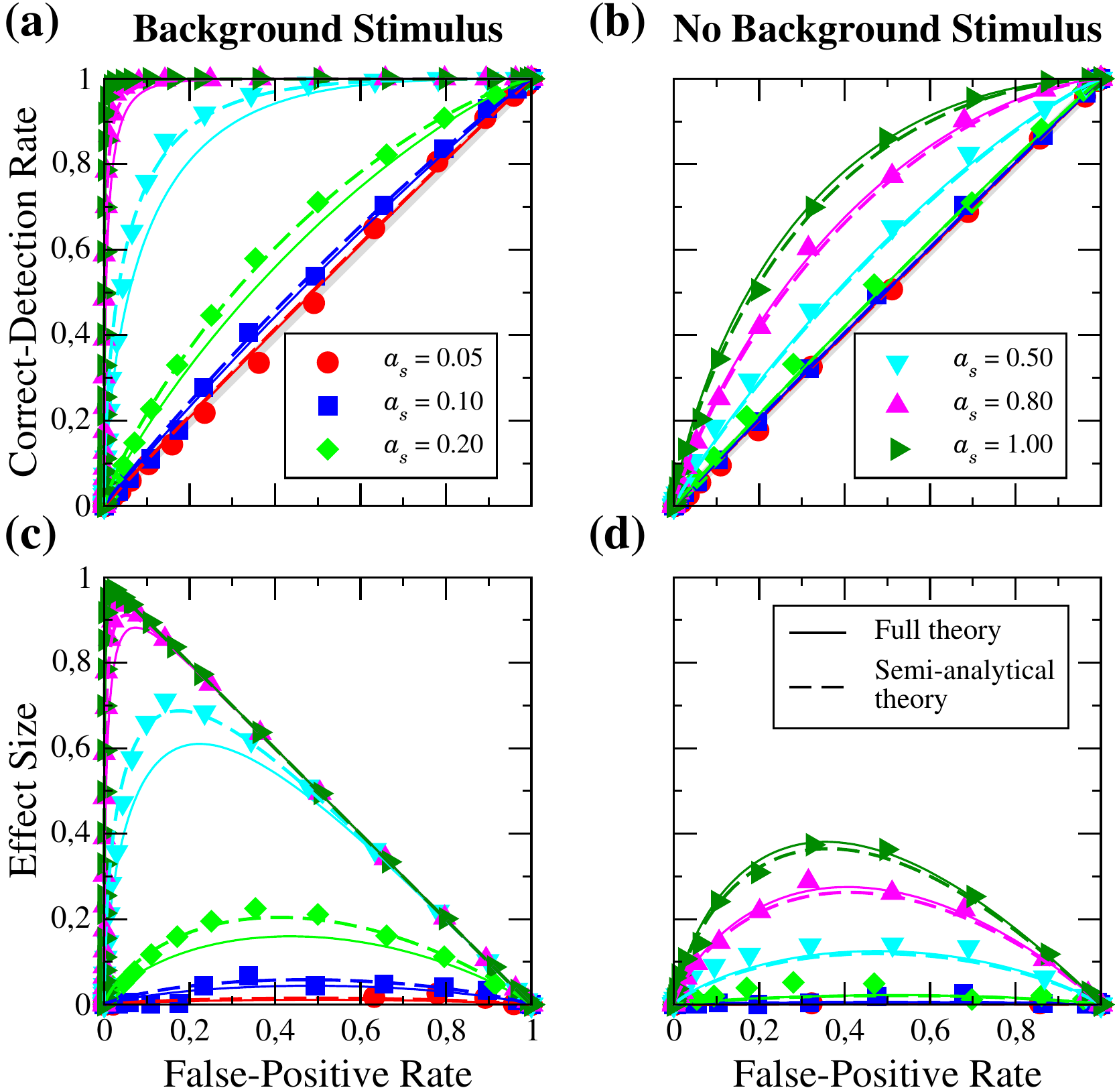}
    \caption{Same as \bi{fig7} but in the mean-driven regime.}
    \label{fig:fig8}
\end{figure}

\begin{figure*}
	\centering
	\includegraphics[width=\textwidth]{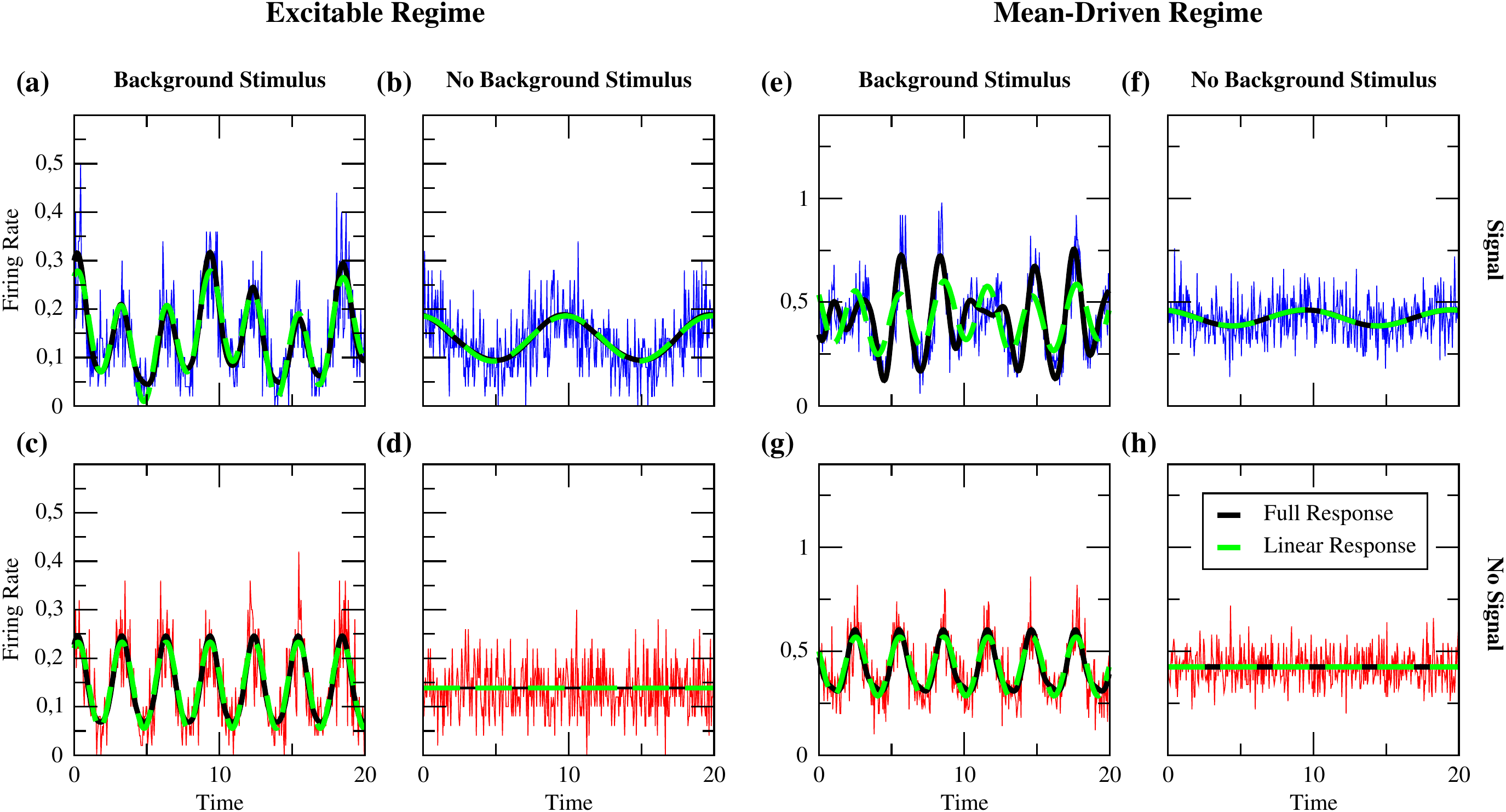}
	\caption{Firing rates ontained from simulations in presence (blue) and absence (red) of the signal  in the excitable (a)-(d) and mean-driven (e)-(h) regime. Black solid lines show the nonlinear response of the firing rate, green dashed lines the linear response.}
	\label{fig:fig9}
\end{figure*}

How does the beneficial effect of the background stimulus  come about? To answer this question we take a closer look at the firing rate in \e{rate_ana}, especially at the influence of response terms of the first (linear response) and second order (higher harmonics and mixed response). \bi{fig9} shows the numerically obtained firing rate with and without the signal , with and without the background stimulus  and in both excitable and deterministic firing regimes. In all panels we also plot the full response up to the second order of the firing rate (solid black lines) as well as the response of the firing rate up to the linear order (dashed green lines).

We find, at least for the chosen frequencies, that the nonlinear response leads to a modestly increased rate modulation in the excitable regime if both background stimulus  and signal  are present (see panel (a)), whereas the nonlinear response has little effect on the rate in the absence of the signal . In marked contrast to this, in the mean-driven regime we observe a strong boost of the firing rate modulation by the nonlinear response (see the pronounced difference between the black and green lines in panel (e)). A stronger rate modulation in the presence of the signal  and the background stimulus  plausibly enhances the detectability of the intruder. Closer inspection of the single contributions of the nonlinear response reveals that the mixed response (MR term in \e{rate_ana}) is responsible for the beneficial boost of the rate modulation.


\subsection{Change of the frequencies of the periodic signals}
So far, we have used a fixed combination of the frequencies of the two periodic signals. Now we investigate how the detectability of the signal  depends on these frequencies. 

For a better visualization we use as a measure of detection performance the \emph{area under the curve} (AUC): For each frequency combination, we measure the area enclosed by the ROC curve and the diagonal (c.f \bi{fig10}). If for instance the ROC curve falls on the diagonal, the performance measure is zero and detection is impossible. Systematic deviations from the diagonal indicate detectability beyond the chance level.

\begin{figure}[h]
	\centering
	\includegraphics[width=0.32\textwidth]{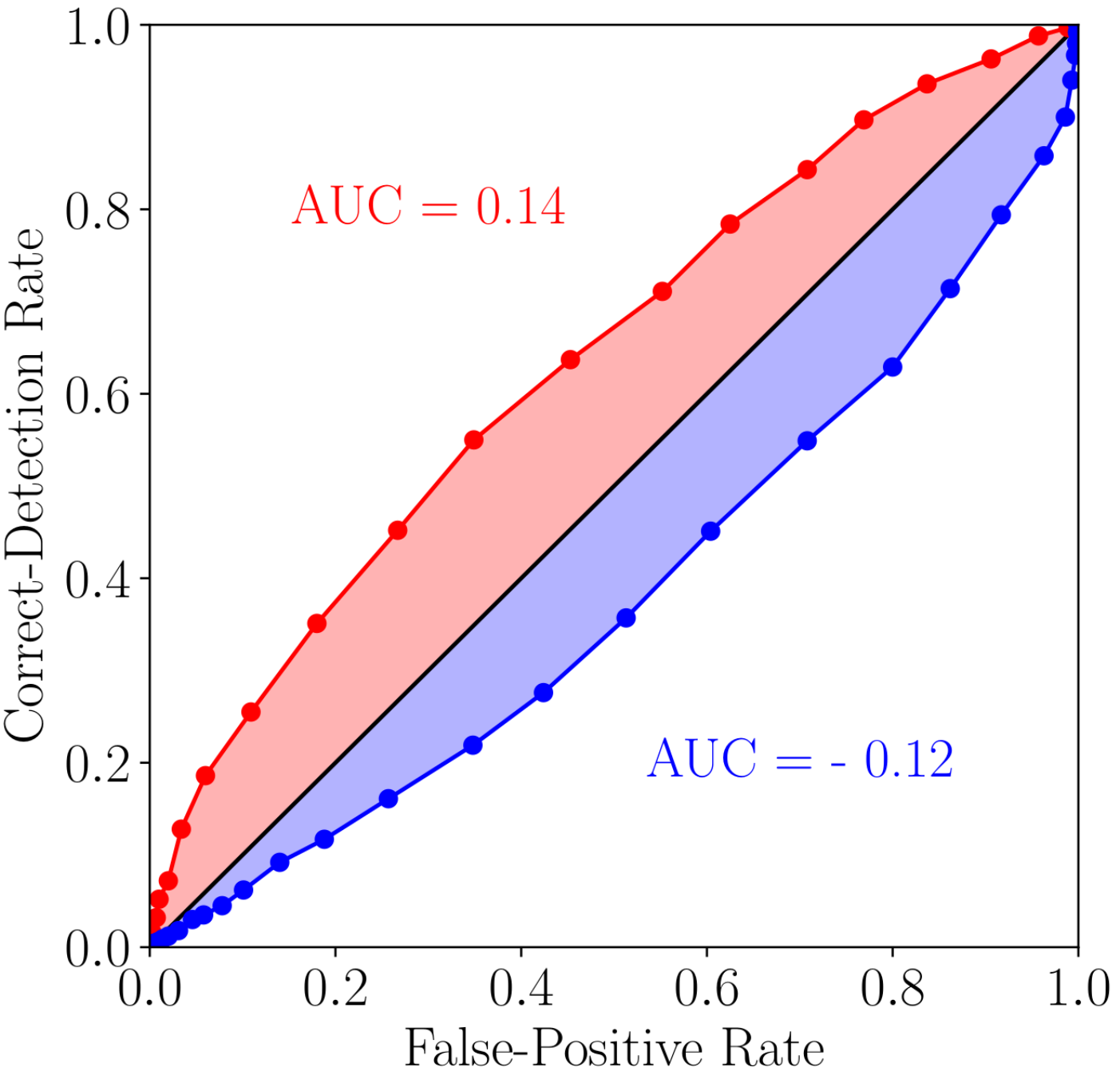}
	\caption{Visualization of the measure of detection perfomance. ROC curves above the diagonal will lead to a positive AUC value, below to a negative one.}
	\label{fig:fig10}
\end{figure}

\begin{figure*}
	\centering
	\includegraphics[width=\textwidth]{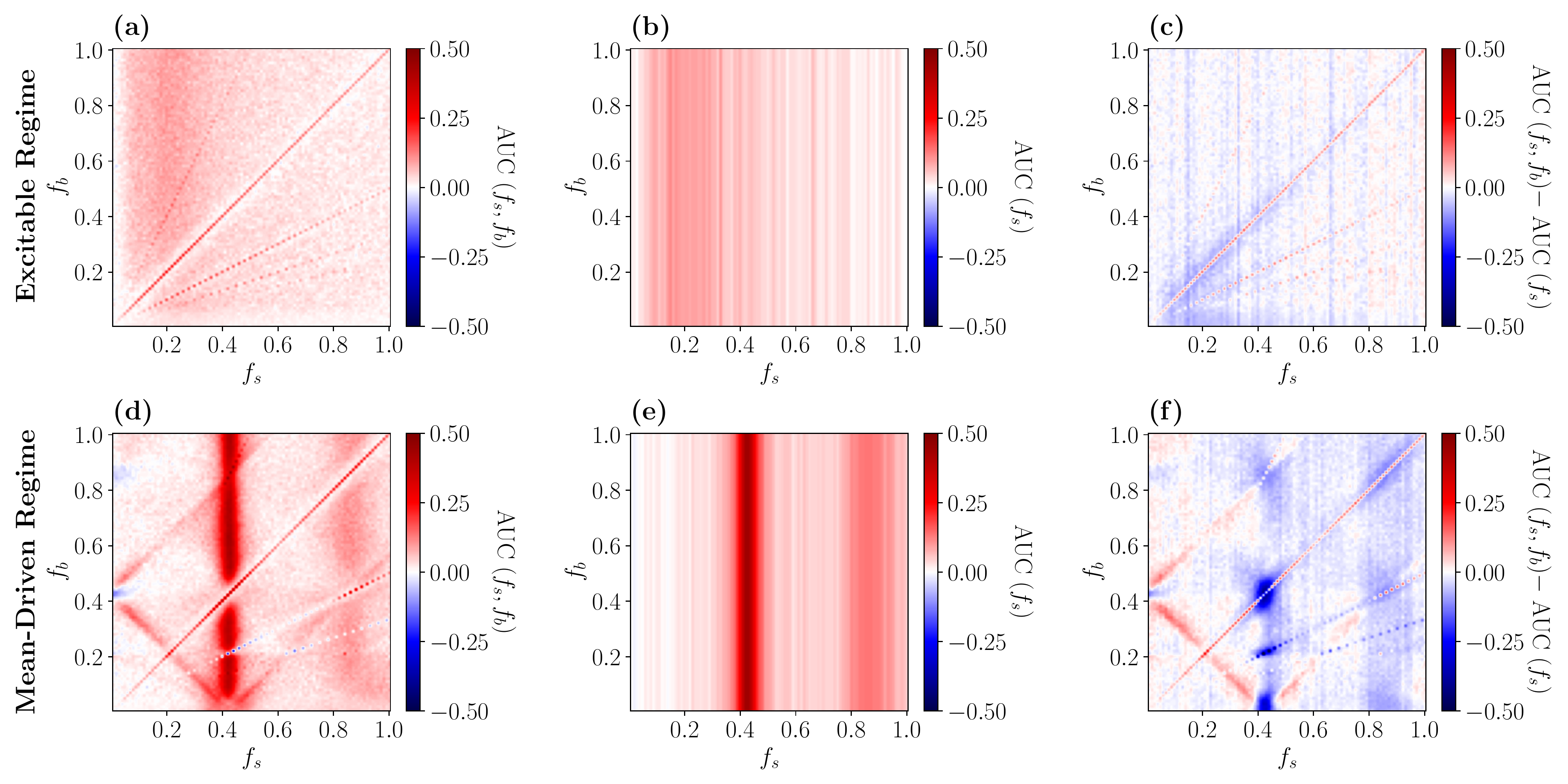}
	\caption{AUC measure in the excitable (a)-(c) and mean-driven regime (d)-(f). (a) and (d) in presence of the female, (b) and (e) in absence of the female, (c) and (f) shows the beneficial effect of the female (red areas). Note, that the colorbar in (c) and (f) is restricted to make the red areas more visible.}
	\label{fig:fig11}
\end{figure*}

The AUC measure allows us to identify conditions under which the presence of the  background stimulus   is beneficial for the detection of the weak periodic signal. We investigate this first for a comparatively weak  signal ($a_{s} = 0.2$) and in the excitable regime, see \bi{fig11}(a)-(c). We subtract the performance in the absence of the  background stimulus   (panel (b)) from the performance in its presence (panel (a)); the difference is shown in panel (c). First of all the detectability seems to be best for a slow  signal (see reddish part on the left in panel (a)); generally, the detectability is not very high in the excitable case for all frequency combinations. It is somewhat better if the background stimulus   is absent (panel (b)) and the difference between the two cases is either close to zero or slightly negative. This coincides with the special case investigated above: for neurons in the excitable regime, the presence of a strong background stimulus   is detrimental for the detection of a faint  signal.

In the mean-driven regime (c.f \bi{fig11}(d)-(f)) the situation becomes more interesting. Turning first to the case where the background stimulus   is absent (see panel (e)), we observe a nearly perfect detection of the signal  for a frequency of $f_{s} = 0.42$ and find a good perfomance for close-by frequencies (note the red vertical stripe around this frequency). The spontaneous firing rate of the neuron population with the used parameters (mean input $\mu$ and noise intensity $D$, see \e{r0}) is $r_{0} \approx 0.42$, so the detectability can be highly improved when the frequency of the driving signal either matches or is close to the spontaneous firing rate of the LIF neurons. Furthermore, the detection performance increases also for $f_{s} \approx 2r_{0}$ but the effect is much weaker (note the fainter red vertical stripe around $f_{s} = 2r_{0}$).

In presence of the background stimulus   (panel (d)) we also have an excellent detection of the weak signal for a signal  frequency of $f_{s} \approx r_{0}$ and for almost all frequencies of the background stimulus   (again, there is a pronounced vertical stripe around this frequency). However, much more striking is the improved detectability on certain diagonal lines on which either $f_{s} + f_{b} \approx r_{0}$ or $\left|f_{s} - f_{b}\right| \approx r_{0}$. These contributions arise due to the weakly nonlinear response of the neurons in the considered dynamical regime. The sum and difference of the two frequencies appear as contributions in the mixed response in \e{rate_ana} and lead to the beneficial role of the  background stimulus   in the detection of the faint signal, which becomes clear when considering the difference of the detectability measure in the presence and absence of the background stimulus   (red areas in panel (f)). We note, that in the motivating example of intruder detection in the courtship situation of weakly electric fish,  the detection benefits for the behaviorely relevant frequency combinations, \ie for $f_{s} < f_{b}$ and for $f_{s} \lessapprox r_{0}$ -- conditions that are a consequence of the distinct distributions of EOD frequencies for male and female weakly electric fish \cite{HenKra18} ($f_{s}$ and $f_{b}$ are the beating frequencies, \ie they result from the difference of EOD frequencies of interacting fish).

Last we want to adress the question, how the detection performance and the nonlinear response, especially the second-order susceptibility $\chi_{2}(f_{1},f_{2})$, are related. For this purpose we use the analytical expression for $\chi_{2}(f_{1},f_{2})$ in combination with the expression for the firing rate modulation and our theory \e{fpr_ana} and \e{cdr_ana} to compute ROC curves for a broad range of frequency combinations $f_{1}, f_{2}$. \bi{fig12} shows the absolute value of the second-order susceptibility (panel (a)) and the AUC measure of the analytical obtained ROC curves (panel (b)).

\begin{figure}[h]
	\centering
	\includegraphics[width=0.48\textwidth]{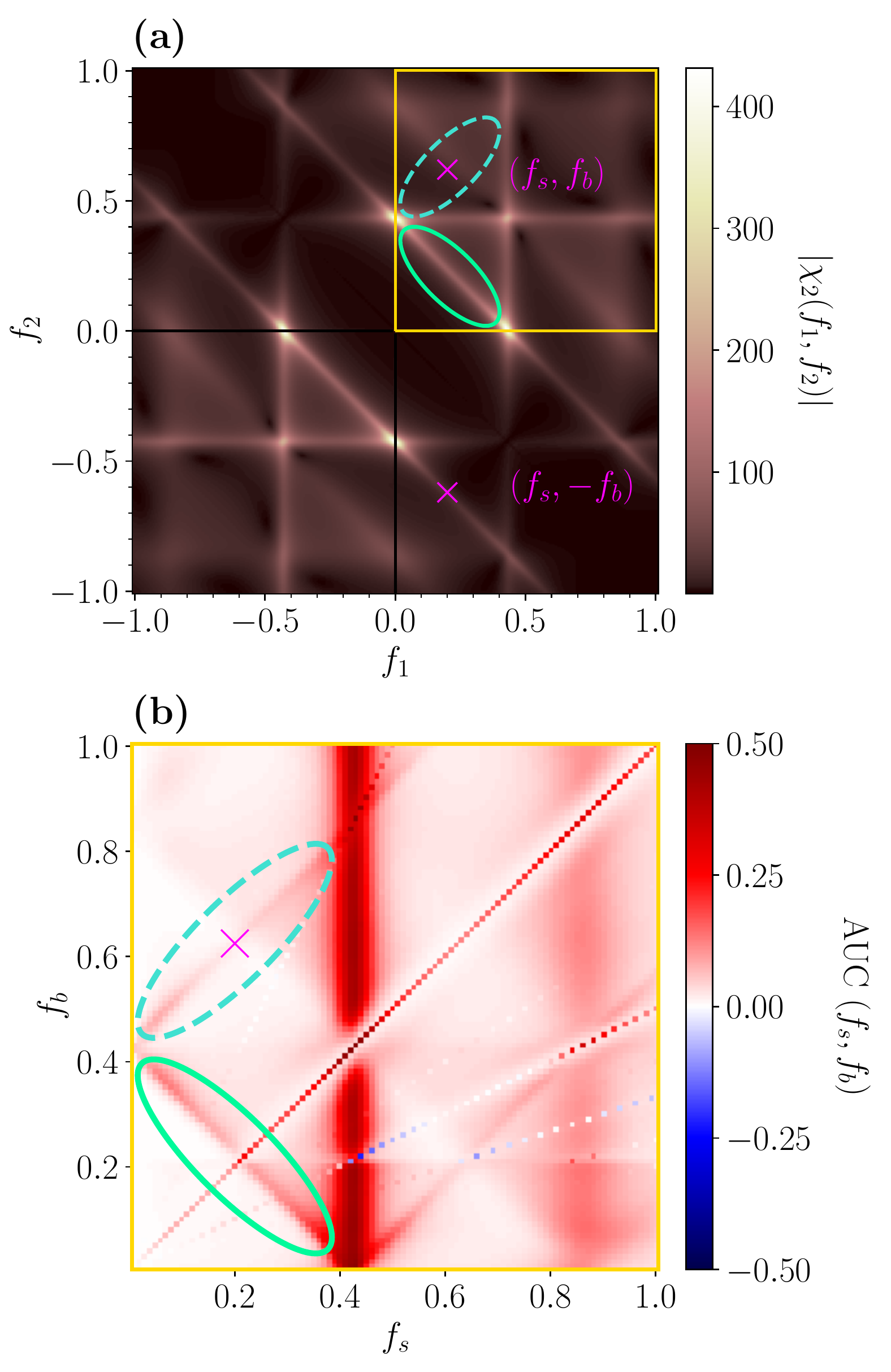}
	\caption{(a) Absolute value of the second-order susceptibility $\chi_{2}(f_{1},f_{2})$, (b) AUC measure in presence of the background stimulus   in the mean-driven regime for ROC curves obtained from weakly nonlinear response theory.}
	\label{fig:fig12}
\end{figure}

Firstly, we observe a very good agreement of the analytically determined AUC measure (\bi{fig12}(b)) with the numerically obtained one (\bi{fig11}(d)). The small deviations of our analytical theory from the simulation results in the ROC curves found in the previous sections result in slightly different AUC values for some frequency combinations, however, the general structure of the detection performance is the same. Secondly, the absolute value of $\chi_{2}(f_{1}, f_{2})$ has some symmetry properties (for a detailed discussion, see \cite{VorLin17}) which the AUC does not share due to the unequal roles of the two frequencies $f_{s}$ and $f_{b}$. Despite this lack of complete symmetry in the AUC there are still similarities but also characteristic differences to $\left|\chi_{2}(f_{1},f_{2})\right|$. The most prominent similarities are the strong maximum around $f_{1} = r_{0}$  ($f_{s} = r_{0}$) and the other maximum around the antidiagonal line (highlighted by a green ellipse in \bi{fig12} both in (a) and (b)) for which $f_{1} + f_{2} = r_{0}$ ($f_{s} + f_{b} = r_{0}$). The most striking differences are that $\left|\chi_2(f_{1}, f_{2})\right|$ is large around the horizontal line $f_{2} = r_{0}$ where the AUC is small and that the AUC is large around the two diagonals on which $\left|f_{s} - f_{b}\right| = r_{0}$ (one of them emphasized by a dashed turquoise ellipse in panel (b)) where the nonlinear response is very small (dashed blue ellipse in panel (a)). The difference between the AUC on the vertical line $f_{s} = r_{0}$ and the horizontal line $f_{b} = r_{0}$ is clear evidence of the unequal roles of the two frequencies. To understand why on the horizontal line the detection performance is not good it suffices to take into account that for $f_{b}$ close to the $r_{0}$, the firing rate is strongly modulated both in presence and absence of the intruder, which makes it harder to detect a faint  signal.

How can we explain the diagonal lines around $\left|f_{s} - f_{b}\right| = r_{0}$ that appear in panel (b) but not in panel (a) for $f_{1}, f_{2} > 0$ ($f_{s}, f_{b} > 0$)? Turning once again to \e{rate_ana}, we note that the mixed response consists of two parts: (i) a term proportional to $\left|\chi_{2}(f_{s}, f_{b})\right|$ -- this contribution is strong when the condition $f_{s} + f_{b} \approx r_{0}$ is fulfilled (antidiagonal highlighted by the green ellipses); (ii) a term proportional to $\left|\chi_{2}(f_{s}, -f_{b})\right|$. The latter is just the mirrored diagonal line around the $f_{1}-$axis of $\left|\chi_{2}(f_{1}, f_{2})\right|$ (compare the dashed turquoise to the solid turquoise ellipse), where the absolute value of the second-order susceptibility is again comparably high. This leads to a strong mixed response not only for frequency combinations around $f_{s} + f_{b} = r_{0}$ but also around $\left|f_{s} - f_{b}\right| = r_{0} $ in the firing rate modulation and thus in the AUC.


\section{Summary and conclusions}

In this paper, we considered a homogeneous population of LIF neurons driven by two periodic stimuli, one playing the role of a strong periodic background and the other one that of a faint signal to be detected. 
We investigated the detection perfomance in the excitable and mean-driven regimes and varied the signal and detection parameters. We developed an analytical framework to calculate ROC curves for the population activity approximately and demonstrated that our formulas which use the linear and the weakly nonlinear response of the instantaneous firing rate work reasonably well.

In general we found in both regimes that the effect of enlarging the detection time window is very weak, whereas increasing the signal  amplitude leads to a strong improvement in the detectability of the faint signal. For specific frequency combinations and in the mean-driven regime, the receiver fish benefits greatly from the presence of the background stimulus : the detectability of the weak signal increases for $a_{b} > 0$. We showed that this effect can be traced back to the weakly nonlinear response and in particular to the so-called mixed response to two periodic signals. In marked contrast, in the excitable case the presence of the background stimulus  was only detrimental to the detection of a faint signal .

With respect to our motivating example, the intruder detection in a courtship situation of weakly electric fish, our modelling assumptions still have severe limitations. Firstly, we assumed a homogenous population of LIF neurons, although encoding populations in the sensory periphery display a pronounced heterogeneity with respect to mean activity (distributions of firing rates are broad) and to variability (also the CV is broadly distributed). For instance, the P-units in weakly electric fish possess firing rates between 50-400 Hz and CV's between 0.2-0.9 \cite{GreKru17}. Furthermore, many neurons in the sensory periphery display a pronounced firing rate adaptation \cite{BenHer03,BenLon05}, that can be incorporated in the integrate-and-fire framework with additional variables (see \eg \cite{LiuWan01,ChaLon01,NauMar08,SchLin15,RamLin21}). It is unclear how neural heterogeneity and adaptation will affect the detectability of a weak stimulus. These are exciting problems for future research.


\begin{acknowledgements}

We would like to acknowledge many helpful discussions with Jan Benda, Jan Grewe and Alexandra Barayeu (University of Tübingen, Germany). We are particularly indebted to Alexandra Barayeu for an intense exchange on the detection performance as a function of the two stimulus frequencies.

\end{acknowledgements}


\appendix

\section{\label{sec:app} Weakly nonlinear response theory}
We briefly state the important measures needed for the applied model (for details see \cite{VorLin17}). Note, that we here consider the LIF model with white noise for a vanishing refractory period, which simplifies the analytical results given in the following somewhat.

The response of the firing rate $r(t)$ to a weak signal $\epi s(t)$ $(\epi \ll 1)$ can be approximated by the first terms of a Volterra series
\begin{align}
	r(t) &= r_{0} + \epi \int \dd{t_{1}'} \ K_{1} (t_{1}') s(t-t_{1}') \nonumber \\
	& \quad + \epi^{2} \int \dd{t_{2}'} \int \dd{t_{2}''} \ K_{2}(t_{2}', t_{2}'') s(t-t_{2}') s(t-t_{2}'') \nonumber \\
	& \quad + \mathcal{O}(\epi^{3}) \ .
	\label{eq:volterra}
\end{align} 
Here, $r_{0}$ is the spontaneous firing rate, \ie if no signal is present ($\epi = 0$), $K_{1}$ and $K_{2}$ are the first-order and second-order response kernels. For the white noise driven LIF model, $r_{0}$ is given by \cite{Ric77}
\begin{align}
	r_{0} = \bqty{\sqrt{\pi} \int_{\frac{\mu - v_{T}}{\sqrt{2D}}}^{\frac{\mu - v_{R}}{\sqrt{2D}}} \dd{x}\ e^{x^{2}} \text{erfc}(x) }^{-1} \ .
	\label{eq:r0}
\end{align}
In the frequency domain \e{volterra} reads
\begin{align}
	\tilde{r}(\omega) &= r_{0} \delta(\omega) + \epi \chi_{1}(\omega) \tilde{s}(\omega) \nonumber \\
	& \quad + \epi^{2} \int \dd{\omega'} \ \chi_{2}(\omega-\omega', \omega') \tilde{s}(\omega-\omega') \tilde{s}(\omega') \nonumber \\
	& \quad + \mathcal{O}(\epi^{3}) \ ,
\end{align}
where $\chi_{1}$ and $\chi_{2}$ are the Fourier transforms of the respective kernels
\begin{align}
	\chi_{1}(\omega) &= \int \dd{t} \ e^{i\omega t)}K_{1}(t) \ , \\
	\chi_{2}(\omega_{1},\omega_{2}) &= \iint \dd{t_{1}} \dd{t_{2}} \ e^{i\omega_{1} t_{1}} e^{i\omega_{2}t_{2}} K_{2}(t_{1},t_{2}) \ .
\end{align}
In terms of the parabolic cylinder functions $\mathcal{D}_{k}(x)$ \cite{AbrSte70} the linear response function $\chi_{1}$ \cite{LinSch01} is given by
\begin{align}
	\chi_{1}(\omega) = \frac{r_{0} i \omega}{\sqrt{D}(i \omega - 1)} \frac{ \D_{i \omega - 1} \pqty{\frac{\mu - v_{T}}{\sqrt{D}}} - e^{\Delta} \D_{i \omega - 1} \pqty{\frac{\mu - v_{R}}{\sqrt{D}}} }{\D_{i \omega} \pqty{\frac{\mu - v_{T}}{\sqrt{D}}} - e^{\Delta}\D_{i \omega} \pqty{\frac{\mu - v_{R}}{\sqrt{D}}}}
\end{align}
with $\Delta = \bqty{v_{R}^{2} - v_{T}^{2} + 2\mu(v_{T} - v_{R})}/(4D)$; an equivalent expression in terms of hypergeometric functions was given by Brunel et al. \cite{BruCha01} (for the equivalence of the expressions, see \cite{Lin02}). The second-order nonlinear response function $\chi_{2}$ has been calculated by Voronenko and Lindner \cite{VorLin17}:
\begin{widetext}
\begin{align}
	\chi_{2}(\omega_{1},\omega_{2}) &= \frac{ r_{0} (1 - i \omega_{1} - i \omega_{2}) (i \omega_{1} + i \omega_{2}) }{2D (i \omega_{1} - 1) (i \omega_{2} -1) } \frac{ \D_{i \omega_{1} + i \omega_{2} - 2} \pqty{\frac{\mu - v_{T}}{\sqrt{D}}} - e^{\Delta} \D_{i \omega_{1} + i \omega_{2} - 2} \pqty{\frac{\mu - v_{R}}{\sqrt{D}}} }{\D_{i \omega_{1} + i \omega_{2}} \pqty{\frac{\mu - v_{T}}{\sqrt{D}}} - e^{\Delta} \D_{i \omega_{1} + i \omega_{2}} \pqty{\frac{\mu - v_{R}}{\sqrt{D}}}} \nonumber \\
	& + \frac{i \omega_{1} + i \omega_{2}}{\sqrt{2D}} \frac{ \pqty{ \frac{\chi_{1} (\omega_{1})}{i \omega_{2} - 1} + \frac{\chi_{1} (\omega_{2})}{i \omega_{1} - 1} } \D_{i \omega_{1} + i \omega_{2} - 1} \pqty{\frac{\mu - v_{T}}{\sqrt{D}}} }{\D_{i \omega_{1} + i \omega_{2}} \pqty{\frac{\mu - v_{T}}{\sqrt{D}}} - e^{\Delta} \D_{i \omega_{1} + i \omega_{2}} \pqty{\frac{\mu - v_{R}}{\sqrt{D}}}} 
	 - \frac{i \omega_{1} + i \omega_{2}}{\sqrt{2D}} \frac{ \pqty{ \frac{\chi_{1} (\omega_{1})}{i \omega_{2} - 1} + \frac{\chi_{1} (\omega_{2})}{i \omega_{1} - 1} } e^{\Delta} \D_{i \omega_{1} + i \omega_{2} - 1} \pqty{\frac{\mu - v_{R}}{\sqrt{D}}} }{\D_{i \omega_{1} + i \omega_{2}} \pqty{\frac{\mu - v_{T}}{\sqrt{D}}} - e^{\Delta} \D_{i \omega_{1} + i \omega_{2}} \pqty{\frac{\mu - v_{R}}{\sqrt{D}}}} \ .
\end{align}
\end{widetext}


\begin{thebibliography}{10}

\bibitem{McD09}
J.~H. McDermott.
\newblock The cocktail party problem.
\newblock {\em Current Biology}, 19:R1024–R1027, 2009.

\bibitem{MarHud01b}
P.~Martin and A.~J. Hudspeth.
\newblock Compressive nonlinearity in the hair bundle's active response to
  mechanical stimulation.
\newblock {\em Proc. Natl. Acad. Sci. USA}, 98:14386, 2001.

\bibitem{FaiSel08}
A.~A. Faisal, L.~P.~J. Selen, and D.~M. Wolpert.
\newblock Noise in the nervous system.
\newblock {\em Nat. Rev. Neurosci.}, 9:292, 2008.

\bibitem{Kni72a}
B.~W. Knight.
\newblock {Dynamics of Encoding in a Population of Neurons}.
\newblock {\em J. Gen. Physiol.}, 59:734, 1972.

\bibitem{Ger00}
W.~Gerstner.
\newblock Population dynamics of spiking neurons: {Fast} transients,
  asynchronous states, and locking.
\newblock {\em Neural Comput.}, 12:43, 2000.

\bibitem{FouHan03}
N.~Fourcaud-Trocm\'e, D.~Hansel, C.~van Vreeswijk, and N.~Brunel.
\newblock How spike generation mechanisms determine the neuronal response to
  fluctuating inputs.
\newblock {\em J. Neurosci.}, 23:11628, 2003.

\bibitem{MazPan08}
A.~Mazzoni, S.~Panzeri, N.~K. Logothetis, and N.~Brunel.
\newblock Encoding of naturalistic stimuli by local field potential spectra in
  networks of excitatory and inhibitory neurons.
\newblock {\em PLoS Comput. Biol.}, 4:e1000239, 2008.

\bibitem{GerKis14}
W.~Gerstner, W.~M. Kistler, R.~Naud, and L.~Paninski.
\newblock {\em Neuronal Dynamics From single neurons to networks and models of
  cognition}.
\newblock Cambridge University Press, Cambridge, 2014.

\bibitem{MusGer19}
S.~P. Muscinelli, W.~Gerstner, and T.~Schwalger.
\newblock How single neuron properties shape chaotic dynamics and signal
  transmission in random neural networks.
\newblock {\em PLoS Comput. Biol.}, 15:e1007122, 2019.

\bibitem{KnoLin22}
G.~Knoll and B.~Lindner.
\newblock Information transmission in recurrent networks: {Consequences} of
  network noise for synchronous and asynchronous signal encoding.
\newblock {\em Phys. Rev. E}, 105:044411, 2022.

\bibitem{HenKra18}
J.~Henninger, R.~Krahe, F.~Kirschbaum, J.~Grewe, and J.~Benda.
\newblock Statistics of natural communication signals observed in the wild
  identify important yet neglected stimulus regimes in weakly electric fish.
\newblock {\em J. Neurosci.}, 38:5456, 2018.

\bibitem{EngZup01}
G.~Engler and G.~Zupanc.
\newblock Differential production of chirping behavior evoked by electrical
  stimulation of the weakly electric fish, apteronotus leptorhynchus.
\newblock {\em J Comp Physiol A}, 187:747, 2001.

\bibitem{NelXu97}
M.~E. Nelson, Z.~Xu, and J.~R. Payne.
\newblock Characterization and modeling of {P-type} electrosensory afferent
  responses to amplitude modulations in a wave-type electric fish.
\newblock {\em J Comp Physiol A}, 181:532, 1997.

\bibitem{VilLin09}
R.~D. Vilela and B.~Lindner.
\newblock Are the input parameters of white-noise-driven integrate \& fire
  neurons uniquely determined by rate and {CV}?
\newblock {\em J. Theor. Biol.}, 257:90, 2009.

\bibitem{BenLon05}
J.~Benda, A.~Longtin, and L.~Maler.
\newblock Spike-frequency adaptation separates transient communication signals
  from background oscillations.
\newblock {\em J. Neurosci.}, 25:2312, 2005.

\bibitem{BerLin20}
D.~Bernardi and B.~Lindner.
\newblock Receiver operating characteristic curves for a simple stochastic
  process that carries a static signal.
\newblock {\em Phys. Rev. E}, 101:062132, 2020.

\bibitem{MidCha03}
J.~W. Middleton, M.~J. Chacron, B.~Lindner, and A.~Longtin.
\newblock Firing statistics of a neuron model driven by long-range correlated
  noise.
\newblock {\em Phys. Rev. E.}, 68:021920, 2003.

\bibitem{Lin06}
B.~Lindner.
\newblock Superposition of many independent spike trains is generally not a
  {Poisson} process.
\newblock {\em Phys. Rev. E.}, 73:022901, 2006.

\bibitem{AbrSte70}
M.~Abramowitz and I.~A. Stegun.
\newblock {\em Handbook of Mathematical Functions}.
\newblock Dover, New York, 1970.

\bibitem{FouBru02}
N.~Fourcaud and N.~Brunel.
\newblock Dynamics of the firing probability of noisy integrate-and-fire
  neurons.
\newblock {\em Neural Comput.}, 14:2057, 2002.

\bibitem{VorLin17}
S.~Voronenko and B.~Lindner.
\newblock {Nonlinear response of noisy neurons}.
\newblock {\em New J. Phys.}, 19:033038, 2017.

\bibitem{HouBre08}
A.~R. Houweling and M.~Brecht.
\newblock Behavioural report of single neuron stimulation in somatosensory
  cortex.
\newblock {\em Nature}, 451:65, 2008.

\bibitem{BerDor21}
D.~Bernardi, G.~Doron, M.~Brecht, and B.~Lindner.
\newblock A network model of the barrel cortex combined with a differentiator
  detector reproduces features of the behavioral response to single-neuron
  stimulation.
\newblock {\em PLoS Comp. Bio.}, 17:e1007831, 2021.

\bibitem{FraRam22}
J.~Franzen, L.~Ramlow, and B.~Lindner.
\newblock The steady state and response to a periodic stimulation of the firing
  rate for a theta neuron with correlated noise.
\newblock {\em J. Comp. Neurosci.}, accepted, 2022.

\bibitem{GreKru17}
J.~Grewe, A.~Kruscha, B.~Lindner, and J.~Benda.
\newblock {Synchronous Spikes are Necessary but not Sufficient for a Synchrony
  Code}.
\newblock {\em PNAS}, 114:E1977, 2017.

\bibitem{BenHer03}
J.~Benda and A.~V.~M. Herz.
\newblock A universal model for spike-frequency adaptation.
\newblock {\em Neural Comput.}, 15:2523, 2003.

\bibitem{LiuWan01}
Y.~H. Liu and X.~J. Wang.
\newblock Spike-frequency adaptation of a generalized leaky integrate-and-fire
  model neuron.
\newblock {\em J. Comput. Neurosci.}, 10:25, 2001.

\bibitem{ChaLon01}
M.~J. Chacron, A.~Longtin, and L.~Maler.
\newblock Negative interspike interval correlations increase the neuronal
  capacity for encoding time-dependent stimuli.
\newblock {\em J. Neurosci.}, 21:5328--5343, 2001.

\bibitem{NauMar08}
R.~Naud, N.~Marcille, C.~Clopath, and W.~Gerstner.
\newblock Firing patterns in the adaptive exponential integrate-and-fire model.
\newblock {\em Biol. Cybern.}, 99:335, 2008.

\bibitem{SchLin15}
T.~Schwalger and B.~Lindner.
\newblock {Analytical approach to an integrate-and-fire model with
  spike-triggered adaptation}.
\newblock {\em Phys. Rev. E Stat. Nonlin. Soft Matter Phys.}, 92:062703, 2015.

\bibitem{RamLin21}
L.~Ramlow and B.~Lindner.
\newblock Interspike interval correlations in neuron models with adaptation and
  correlated noise.
\newblock {\em PLoS Comput. Biol.}, 17:e1009261, 2021.

\bibitem{Ric77}
L.~M. Ricciardi.
\newblock {\em Diffusion Processes and Related Topics on Biology}.
\newblock Springer-Verlag, Berlin, 1977.

\bibitem{LinSch01}
B.~Lindner and L.~Schimansky-Geier.
\newblock Transmission of noise coded versus additive signals through a
  neuronal ensemble.
\newblock {\em Phys. Rev. Lett.}, 86:2934–7, 2001.

\bibitem{BruCha01}
N.~Brunel, F.~S. Chance, N.~Fourcaud, and L.~F. Abbott.
\newblock Effects of synaptic noise and filtering on the frequency response of
  spiking neurons.
\newblock {\em Phys. Rev. Lett.}, 86:2186, 2001.

\bibitem{Lin02}
B.~Lindner.
\newblock {\em Coherence and Stochastic Resonance in Nonlinear Dynamical
  Systems}.
\newblock Logos-Verlag, Berlin, 2002.

\end{thebibliography}
\end{document}